\documentclass[aps,prx,twocolumn,reprint,groupedaddress,amsmath,amssymb,longbibliography]{revtex4-1}
\usepackage{graphicx}  
\usepackage{dcolumn}   
\usepackage{bm}        
\usepackage{verbatim}   
\usepackage{subfigure}
\usepackage{amsmath}
\usepackage{natbib}
\usepackage{xcolor}
\usepackage{lipsum} 
\usepackage{hyperref}

\renewcommand{\Re}{\operatorname{\mathrm{Re}}}

\newcommand{\CV}{\operatorname{\mathrm{CV}}}
\newcommand{\erf}{\operatorname{\mathrm{erf}}}

\begin{document}
	
\title{Solving the 2-Dimensional Fokker-Planck Equation for Strongly Correlated Neurons}
\author{Ta\c{s}k{\i}n Deniz \& Stefan Rotter}
\affiliation{
   Bernstein Center Freiburg \& Faculty of Biology,
   University of Freiburg,
   Hansastra{\ss}e 9a, 79104 Freiburg, Germany
   }
\date{\today}
\begin{abstract}  
Pairs of neurons in brain networks often share much of the input they receive from other neurons. Due to essential non-linearities of the neuronal dynamics, the consequences for the correlation of the output spike trains are generally not well understood. Here we analyze the case of two leaky integrate-and-fire neurons using a novel non-perturbative approach. Our treatment covers both weakly and strongly correlated dynamics, generalizing previous results based on linear response theory. 
 \end{abstract}

\maketitle

\section{Introduction.}Both membrane potentials and action potentials recorded from nearby neurons in networks of the brain exhibit non-trivial statistical dependencies, typically quantified by cross correlation functions \cite{Lampl1999,Okun2008,Poulet2008}. Theoretical models have emphasized that such correlations are an inevitable consequence if two neurons are part of the same network and share some synaptic input \cite{Ostojic2009,Pernice2011,Pernice2012}. However, for non-linear neuron models, correlation functions are difficult to compute explicitly, especially for low firing rates in the strongly correlated regime \cite{Rosenbaum2012, Schultze-Kraft2013}. Previous analytical approaches have employed perturbation theory \cite{Brunel1999, Lindner2001} to study pair correlations under the assumption of weak input correlation \cite{DelaRocha2007, Shea-Brown2008}. However, there is ample evidence of massive shared input for pairs of nearby neurons, resulting in strong correlations particularly of their membrane potentials \cite{Lampl1999,Okun2008,Poulet2008}. A full theory of correlations, covering the case of both weak and strong shared input alike, demands non-perturbative methods that take non-linear effects into account \cite{Schultze-Kraft2013}.  In the work presented here, we suggest a non-perturbative solution to the corresponding two-dimensional Fokker-Planck equation to describe correlated integrate-and-fire neurons in any regime, with arbitrary precision. We demonstrate that our theoretical predictions accurately fit to correlation functions computed from simulated spike trains. 

Similar problems were studied analytically for arbitrary input correlations of the stochastic dynamics of neural oscillators \cite{Abouzeid2011} and for level-crossings of correlated Gaussian processes \cite{Tchumatchenko2010a}.  Related numerical work considered strong input correlations for integrate-and-fire neurons receiving white noise input \cite{Vilela2009} or receiving shot noise input with nontrivial temporal correlations \cite{Schwalger2015, Voronenko2015}. Additionally, the problem of how to calculate the stationary distributions conditional on a spike from the exit current at the threshold is also discussed in the case of colored noise \cite{Schwalger2015}. Our study further suggests a novel technique to solve 2D Fokker Planck equations for leaky integrate-and-fire neurons, which provides the accurate  steady state joint distribution of membrane potentials. 

\section{Model and Theory.} We consider two leaky integrate-and-fire (LIF) model neurons receiving correlated inputs. Their dynamics are governed by the following stochastic differential equations 
\begin{eqnarray}
\label{eq_langevin_2D}
\tau_a\dot{V_a}=-V_a+\tau_a(\mu_a + \sigma_a [\sqrt{1-c}\:\xi_a\pm\sqrt{c}\:\xi_c])
\end{eqnarray}
where input $I_a=\mu_a + \sigma_a [\sqrt{1-c}\:\xi_a\pm\sqrt{c}\:\xi_c]$ 
with private white noise $\xi_a $ ($a= 1,2$) and shared white noise
$\xi_c$, all components being independent. Input correlation
coefficient is given as $\pm c$, where $0 \leq c < 1$ and $\tau_a$,
$\mu_a$ and $\sigma_a$  are constant parameters characterizing both
the neuron model and the input. Without loss of generality we take only the positive sign in $\pm\sqrt{c}$. 
We parametrize the input  by 
	\begin{eqnarray}
	\label{sup_eq_mu_and_sigma}
	\mu_{a}=J_{Ea} \nu_{Ea}-J_{Ia} \nu_{Ia}\\
	\sigma_{a}=\sqrt{J_{Ea} ^2 \nu_{Ea}+J_{Ia} ^2 \nu_{Ia}} 
	\end{eqnarray}
	where $J_{Ea} $ and $J_{Ia}$ represent the amplitude of postsynaptic potentials for excitatory and inhibitory input spike trains. We distinguish input parameters ($J_{Ia}$, $J_{Ia}$, $\nu_{Ea}$, $\nu_{Ia}$) from intrinsic parameters ($\tau_a$, $V_{ra}$, $V_{ta}$).
	
The corresponding  Fokker-Planck equation is 
\begin{widetext}
\begin{equation}
\label{eq_fokker_planck_2D}
\frac{\partial P}{\partial t} = \partial_1 \Big{(}\Big{(}\frac{V_1}{\tau_1}-\mu_1\Big{)}P\Big{)}+\partial_2 \Big{(}\Big{(}\frac{V_2}{\tau_2}-\mu_2\Big{)}P\Big{)}+
\frac{1}{2}
\begin{pmatrix}
\partial_1  & \partial_2  \\
\end{pmatrix}  
\begin{pmatrix}
\sigma_1^2& c\sigma_1 \sigma_2\\
c\sigma_1 \sigma_2 & \sigma_2^2
\end{pmatrix}
\begin{pmatrix}
\partial_1  \\ \partial_2  
\end{pmatrix}  P
\end{equation}
where we define $\partial_a\equiv \frac{\partial}{\partial V_a}$ and $P\equiv P(V_1,V_2,t)$. Using the new variables $x=\frac{V_1-\mu_1\tau_1}{\sigma_1\sqrt{\tau_1}}$ and  $y=\frac{V_2-\mu_2\tau_2}{\sigma_2\sqrt{\tau_2}}$ the equation can be rewritten as 
\begin{align}
\label{eq_fp2d}
\frac{\partial P}{\partial t} &=\frac{1}{\tau_1} \mathcal{L}_1 P +\frac{1}{\tau_2} \mathcal{L}_2 P +\frac{c}{\sqrt{\tau_1\tau_2}} \mathcal{L}_{12} P\\
\mathcal{L}_1P &= \frac{\partial (xP)}{\partial x}+\frac{1}{2}\frac{\partial^2 P}{\partial x^2} \\
\mathcal{L}_2P &= \frac{\partial (yP)}{\partial y}+ \frac{1}{2}\frac{\partial^2 P}{\partial y^2}\\
\mathcal{L}_{12} &= \frac{\partial^2 P}{\partial x \partial y}.
\end{align}
\end{widetext}
The first two terms with operators $\mathcal{L}_1$ and $\mathcal{L}_2$ represent independent populations, and 
they fully describe the 2D dynamics for $c=0$. The third term
represents the correlated diffusion for $c>0$. 

\begin{figure}
	\includegraphics[width=8.6cm]{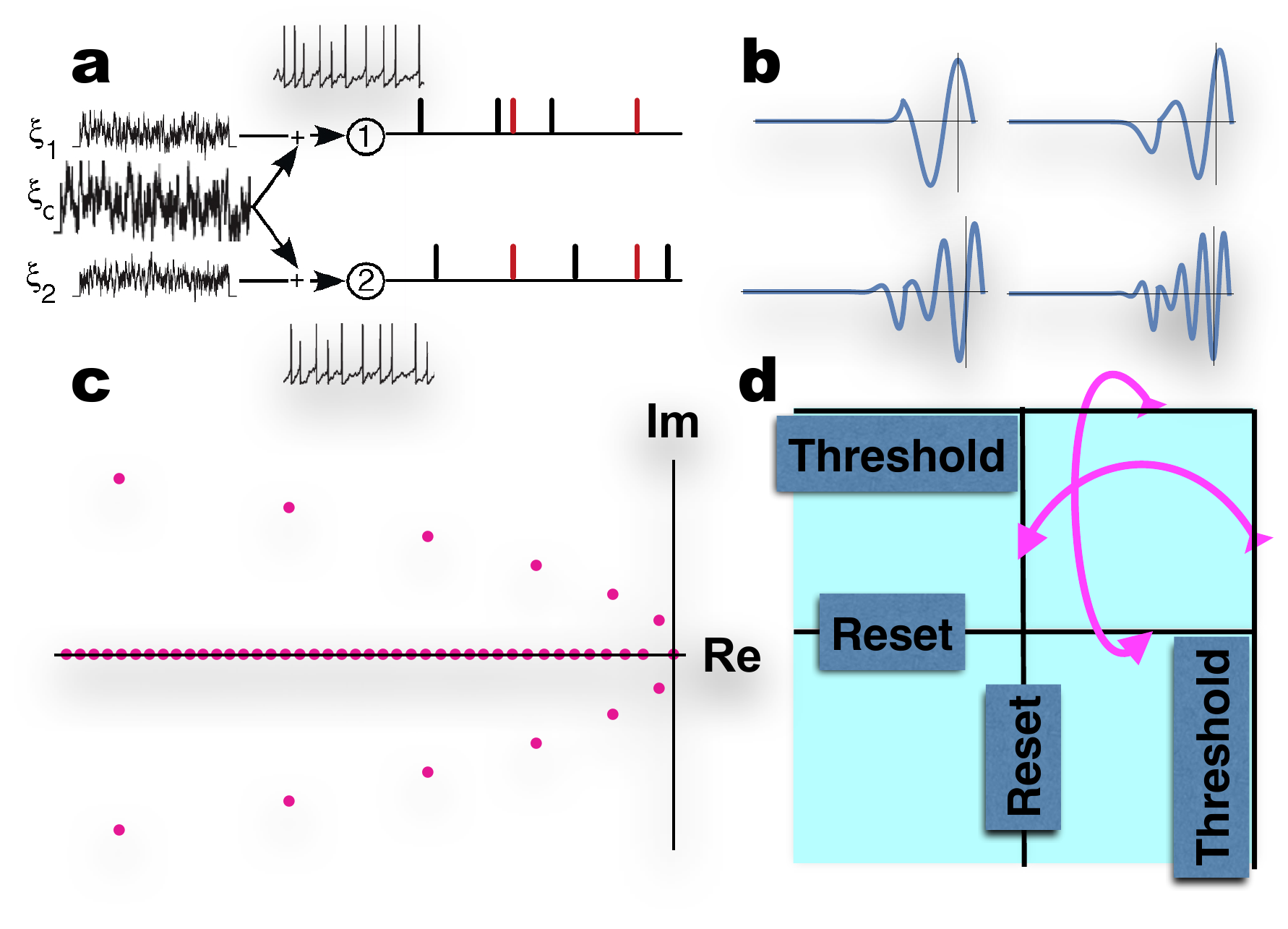}
	\caption{\label{fig1} (a)~Leaky integrate-and-fire neurons driven by strong shared noise, inducing synchronicity in the output spike trains. (b)~Examples of eigenfunctions with increasing $|\Re(\lambda)|$. (c)~A typical discrete eigenvalue spectrum of the diffusion-based LIF model, comprising both real and complex conjugate pairs of eigenvalues with $\Re(\lambda)\leqslant0$. (d)~ Boundary conditions in 2D voltage space with threshold potentials $x_t$, $y_t$ and resting potentials $x_r$, $y_r$.  The magenta arrows represent the reset mechanism once the threshold was hit and a spike was elicited in either neuron. }
\end{figure}

In order to calculate the cross-covariance function of output spike trains,
we first compute the joint steady state distribution of membrane
potentials from
	\begin{equation}
	\label{eq_2d_stationary}
	0 = \frac{1}{\tau_1} \mathcal{L}_1 P_0 +\frac{1}{\tau_2} \mathcal{L}_2 P_0 +\frac{c}{\sqrt{\tau_1\tau_2}} \mathcal{L}_{12} P_0.
	\end{equation} 
We have threshold potentials $x_t$, $y_t$, reset potentials $x_r$, $y_r$ and boundary conditions
\begin{subequations}
	\label{eq_bc_2D}
\begin{eqnarray}
	 P_0(x,y_{t})= 0=P_0(x_{t},y) \qquad\\
	 P_0(x,-\infty)= 0=P_0(-\infty,y) \qquad\\
	 \partial_xP_0(x_{r}-\epsilon,y)-\partial_xP_0(x_{r}+\epsilon,y)\stackrel{\epsilon \rightarrow 0}{= }\partial_xP_0(x_{t},y) \qquad \\
	 \partial_yP_0(x,y_{r}-\epsilon)-\partial_xP_0(x,y_{r}+\epsilon)\stackrel{\epsilon \rightarrow 0}{= }\partial_xP_0(x,y_{t}) \qquad 
	\end{eqnarray} 
\end{subequations}
We derive an expansion of the stationary equation in terms of
eigenfunctions of the uncoupled operators (See Appendix for details.)
$\mathcal{L}_1$ and $\mathcal{L}_2$,
 \begin{equation}
 \label{eq_1d_eigenspace_1}
\mathcal{L}_1 f_{i} = \lambda_{1i} f_{i} \: , \quad \mathcal{L}_2 g_i = \lambda_{2i} g_i 
 \end{equation}
  with boundary conditions given as
  \begin{align}
  \label{eq_bc_1}
  f_i(x_{t})=& \: 0=\lim_{x\rightarrow -\infty} f_i(x) \\
  \partial_x f_i(x_{t})&\stackrel{\epsilon \rightarrow 0}{= } \partial_x f_i(x_r-\epsilon)-\partial_x f_i(x_r+\epsilon).
  \end{align}
Analogous expressions hold for $g_i(y)$. The eigenvalue spectrum of this problem is countable with both real and pairs of complex conjugate eigenvalues (Fig.~\ref{fig1}c). (We assume here that the index $i$ increases with $|\Re(\lambda_i)|$.) In order to expand the solution in the eigenspace of a non-selfadjoint differential operator, the dual eigenvalue problem needs to solved as well (see  Appendix for details.)
  \begin{eqnarray}
  \mathcal{L}_1^{\dagger} \tilde{f}_i &= \lambda_{1i} \tilde{f}_i \: , \quad \mathcal{L}_2^{\dagger} \tilde{g}_i &= \lambda_{2i} \tilde{g}_i
  \end{eqnarray}
  with conjugate boundary conditions 
  \begin{eqnarray}
  \tilde{f}_i(x_{t}) &= \tilde{f}_i(x_r) \: , \quad\tilde{g}_i(y_{t}) &= \tilde{g}_i(y_r).  
  \end{eqnarray}  
This guarantees that the basis $\{ f_i \}$ and the conjugate basis $\{ \tilde{f}_i \}$ are bi-orthogonal in Hilbert Space
\begin{equation}
\int_{-\infty}^{x_{t}} \tilde{f}_i(x) f_j(x)\,dx = \delta_{ij}
\end{equation} 

where we select free coefficients to satisfy bi-orthonormality. 
The solution to Eq.~\ref{eq_fp2d} can now be expanded in terms of functions that individually satisfy the boundary conditions Eq.~\ref{eq_bc_2D}
\begin{equation}
\label{eq_series_expansion_2d}
P_0(x,y) = f_0(x) g_0(y) + F(x) S G(y)
\end{equation}
where we define $F(x) S G(y) \equiv \sum_{ij} S_{ij} f_i(x) g_j(y)$, for some coefficients $S_{ij} \in \mathbb{C}$.
This expansion exactly satisfies the constraints for marginal distributions
\begin{equation}
\int_{-\infty}^{y_{t}} P_0(x,y)\,dy = f_0(x) , \; \int_{-\infty}^{x_{t}} P_0(x,y)\,dx = g_0(y)
\end{equation}
where the probability density function $f_0$ is given by
\begin{eqnarray}
\label{eq_marginal_dist}
f_0(x) = 2r_1\tau_1e^{-x^2} \int_{x}^{x_{t}}\Theta(u-x_r) e^{u^2}\,du,
\end{eqnarray}
where $\Theta(x)$ is the Heaviside step function. The density $g_0(y)$ is defined analogously. 
Steady state firing rates of both neurons are given by
\begin{equation}
\label{eq_steady_state_rate}
r_1 = \frac{1}{\tau_1}\Bigl[\int_{0}^{\infty} e^{-u^2} \frac{e^{x_{t} u}-e^{x_{r} u}}{u}\,du\Bigr]^{-1}
\end{equation}
and a similar expression for $r_2$. Using Eq.~\ref{eq_1d_eigenspace_1}, the solution can now implicitly be written in terms of eigenfunctions
\begin{multline}
\label{eq_2d_in_basis}
F(x) \Lambda_1 S G(y) + F(x) S \Lambda_2 G(y) +\tilde{c}\partial_x F(x) S \partial_y G(y) \\ = -\tilde{c}\partial_x f_0(x) \partial_y g_0(y)
\end{multline}
with diagonal matrix $\Lambda_{a,ij} = \frac{\lambda_{ai} \delta_{ij}}{\tau_a}$ and constant $\tilde{c}=\frac{c}{\sqrt{\tau_1\tau_2}}$. In order to actually solve Eq.~\ref{eq_2d_in_basis} we express the action of the derivative operators on the eigenbasis as
 \begin{align}
 X_{ij} &= \int_{-\infty}^{x_{t}} \tilde{f}_i(x) \partial_xf_j(x)\,dx  
 \end{align}
and similarly for $Y$. The final equation in matrix form is 
\begin{equation}
\label{eq_S_matrix}
\Lambda_1 S +  S \Lambda_2  +\tilde{c} X^{T}S Y = - \tilde{c}X_0 \otimes Y_0 \; .
\end{equation} 
\begin{figure}
	\includegraphics[width=8.6cm]{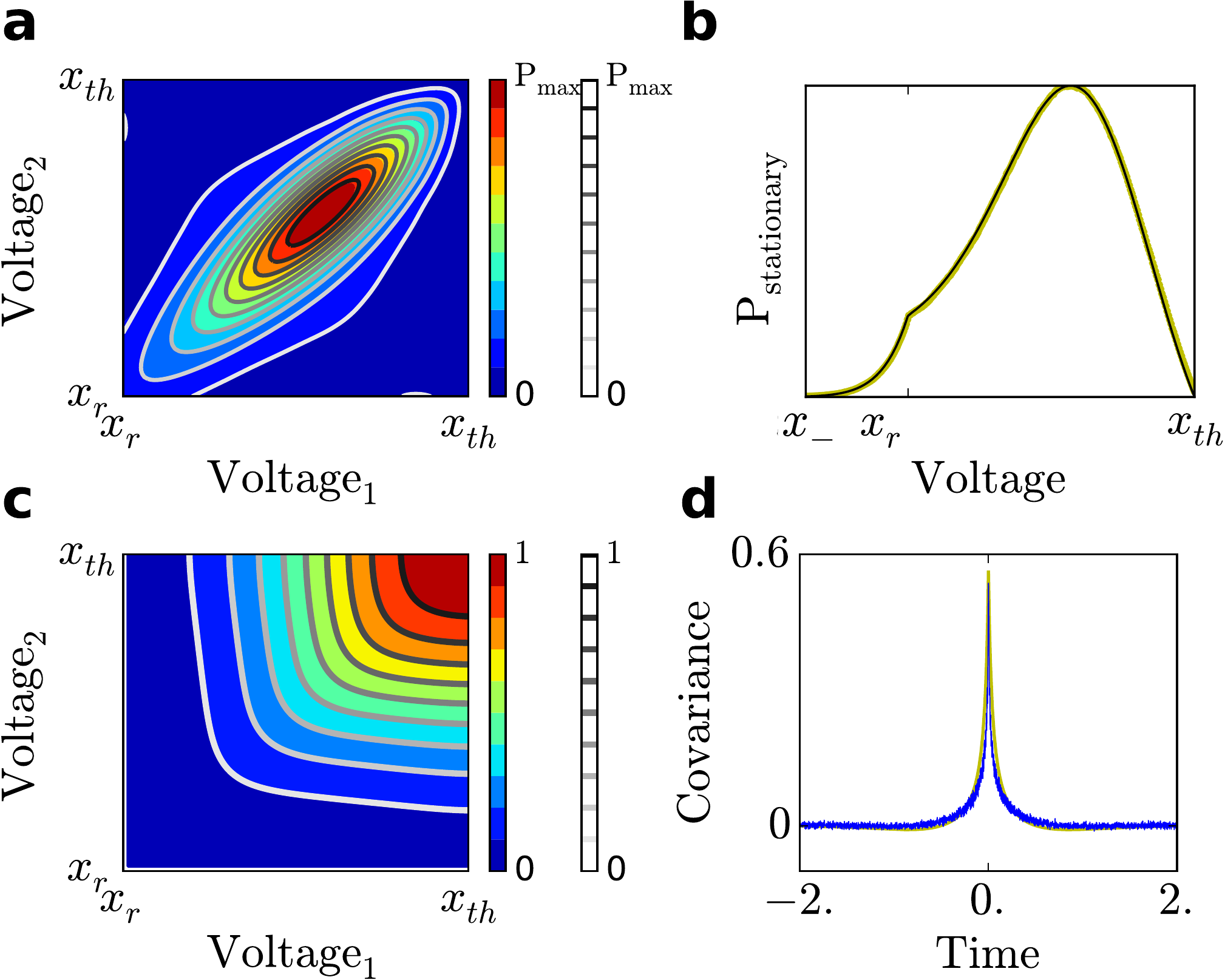}
	\caption{\label{fig2} Simulations and theory yield practically identical results, demonstrated here for $x_r=y_r=-2.0$, $x_t=y_t=0.8$ and $c=0.9$. (a)~Joint membrane potential distribution of simulated data (smoothed 2D histogram of simultaneously recorded membrane potentials), compared to $P_0(x,y)$, Eq.~\ref{eq_2d_in_basis}. The $L_1$ error is approx.~$0.02$, partially caused by a boundary effect for discrete-time simulations of Eq.~\ref{eq_langevin_2D}. (b)~Marginal distribution $f_0(x)$, Eq.~\ref{eq_marginal_dist} (black: data, yellow: theory). (c)~Same as (a), comparison between cumulative distributions $\int_{-\infty}^{x} \int_{-\infty}^{y} P_0(x,y) \,dx dy$.  (d)~Symmetric correlation function $C_{12}(\tau)$ with time rescaled by $\tau_1$. The blue curve is the covariance function of simulated spike trains, while the yellow curve is a numerical rendering of the theory developed here.}
\end{figure}
\section{Spike Train Correlations.}The covariance function of two stationary spike trains $\mathbb{S}_a(t) = \sum_l \delta(t-t^a_l)$ ($a = 1,2$) is given as
\begin{equation}
C_{12}(\tau)= \langle \mathbb{S}_1(t) \mathbb{S}_2(t) \rangle 
-\langle \mathbb{S}_1(t)\rangle \langle \mathbb{S}_2(t) \rangle 
\end{equation}
where $\langle \mathbb{S}_a(t)\rangle = r_a$, with $\langle . \rangle$ indicating the ensemble average.
Using renewal theory, it can be expressed in terms of the conditional firing rate $r_{1|2}(\tau)$ as
\begin{equation}
\label{eq_corr_rate}
C_{12} (\tau)= r_2 (r_{1|2}(\tau)-r_1).
\end{equation}
We derive the conditional firing rate from the stationary joint membrane potential distribution $P_0(x,y)$
via the distribution of the membrane potential conditional to a spike at $t_0 = 0$ found as $P_{1|2}(x)=-\frac{1}{2 r_2\tau_2} \partial_{y} P_0(x,y_{t})$, since $\int_{-\infty}^{x_{t}} \partial_{y} P_0(x,y_{t})\,dx = -2 r_2\tau_2$ by construction. Therefore, we have to solve the initial value problem
\begin{align}
f(t_0,x) &= -\frac{1}{2 r_2\tau_2} \partial_{y} P_0(x,y_{t})\\
\tau_1\partial_t f &= \mathcal{L}_1 f
\end{align}
where $\mathcal{L}_1$ is the time evolution operator in Eq.~\ref{eq_1d_eigenspace_1}. The instantaneous conditional rate in Eq.~\ref{eq_corr_rate} is then $r_{1|2}(t) = -\frac{1}{2\tau_1} \partial_{x} f(x,t)$. The instantaneous conditional distribution is given by 
\begin{equation}
\label{eq_conditional_distribution}
f(x,t) = f_0(x) + \frac{1}{2 r_2\tau_2} \sum_i \Bigl(\sum_j S_{ij} \Bigr) e^{\lambda_{1i} t/\tau_1} f_{i}(x).
\end{equation}
The exit flux at threshold $r_{1|2}(t)$ inserted into Eq.~\ref{eq_corr_rate}
yields the covariance function
\begin{equation}
\label{eq_cov_func}
C_{12}(\tau) = \frac{1}{4\tau_1\tau_2}\sum_{ij} [\Theta(\tau) e^{\Lambda_1 \tau} S + \Theta(-\tau)  S e^{-\Lambda_2 \tau}]_{ij} 
\end{equation}
for $\tau = t_1-t_2$ and $\Lambda_{a,ij} = \frac{\lambda_{ai} \delta_{ij}}{\tau_a}$. Using the symmetry $C_{12}(\tau) = C_{21}(-\tau)$ we obtain the covariance function for negative time lags as well. The correlation coefficient  as considered in \cite{Shea-Brown2008} is computed as (see Appendix for details)
\begin{equation}
\label{eq_corr_coeff}
 C_{out}(c)=\frac{-\tilde{c}}{4{CV_1 CV_2\sqrt{r_1r_2}} }\sum_{ij}\frac{( X S Y+X_0Y_0)_{ij}}{ \Lambda_{1,i}\Lambda_{2,j}}
 \end{equation}
with $CV_a$ being the coefficients of variation of the two output spike trains. Here one can see how the correlation transfer depends non-linearly on $c$ as $S$ is a non-linear function of $c$. 

\begin{figure}
	\includegraphics[width=8.6cm]{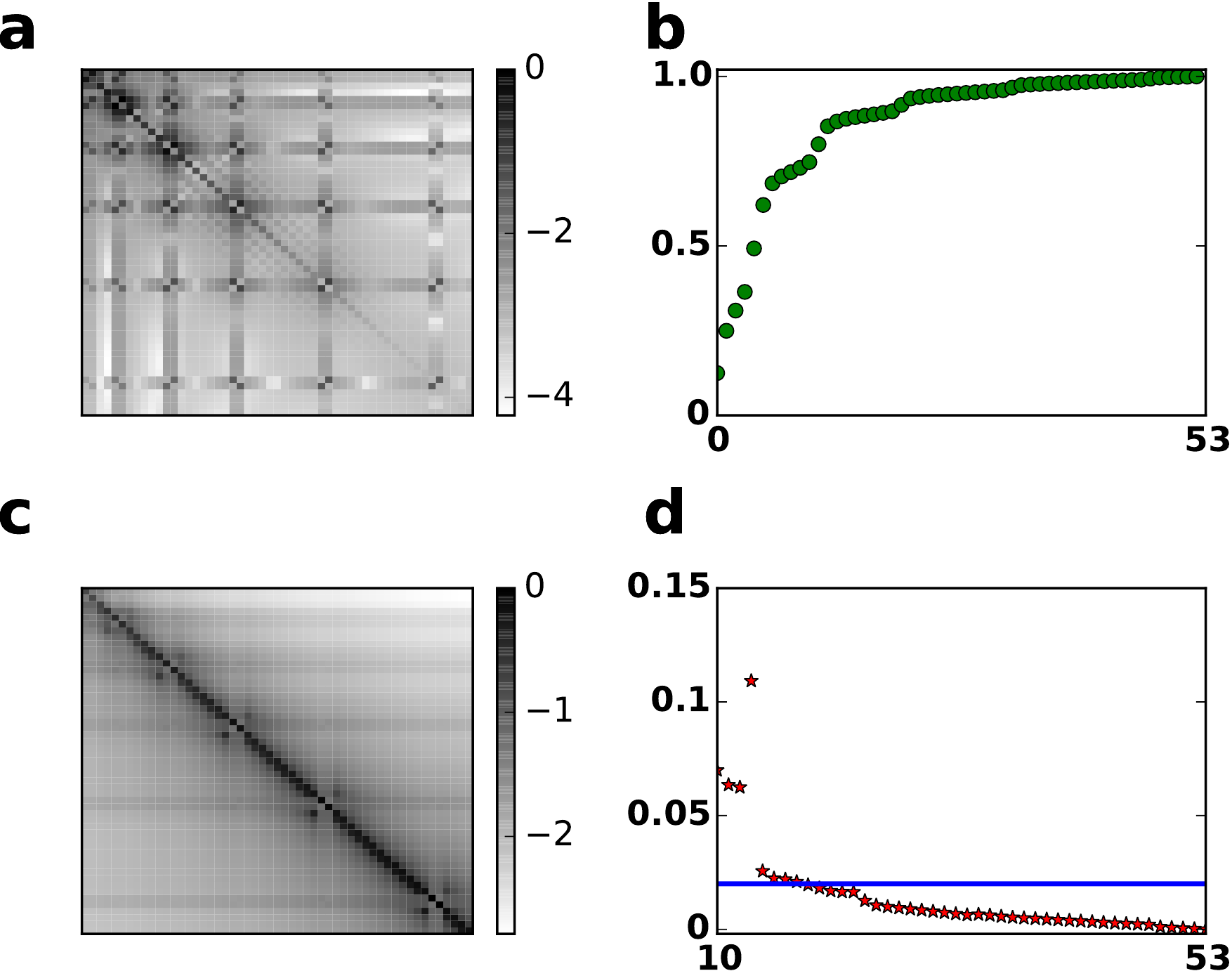}
	\caption{\label{fig3} Numerical example solution with $x_r=y_r=-2.0$, $x_t=y_t=0.8$ and  $c=0.9$. (a)~logarithmic rendering $\log_{10}(|S_{ij}|/\max(|S_{ij}|))$ of mode coupling matrix $S$ ( size: $53 \times 53$).  (b)~Relative convergence of correlation coefficients $C^{(\mathrm{rel})}_n=\sum_{j=1}^n\sum_{i}(S\Lambda^{-1})_{ij}/\sum_{ij}(S\Lambda^{-1})_{ij}$, in Eq.~\ref{eq_corr_coeff}. (c)~Matrix representation $X$ of the derivative operators presented as in (a).  (d)~Relative error $ \int_{-\infty}^{x_r}\int_{-\infty}^{y_r} dx dy |P_0^{N}(x,y)-P_0^{(n)}(x,y)| $, where $n$ is truncation number and $N=53$ is the maximum truncation number in Eq.~\ref{eq_series_expansion_2d}. $N$ is the number of eigenvalues with property $|Re(\lambda_i)| < 100 $. Here we solved Eq.~\ref{eq_S_matrix} for different $n$. The blue line is the $L_1$ error in Fig.~\ref{fig2}a.}
\end{figure}

	\begin{figure}
		\includegraphics[width=8.6cm]{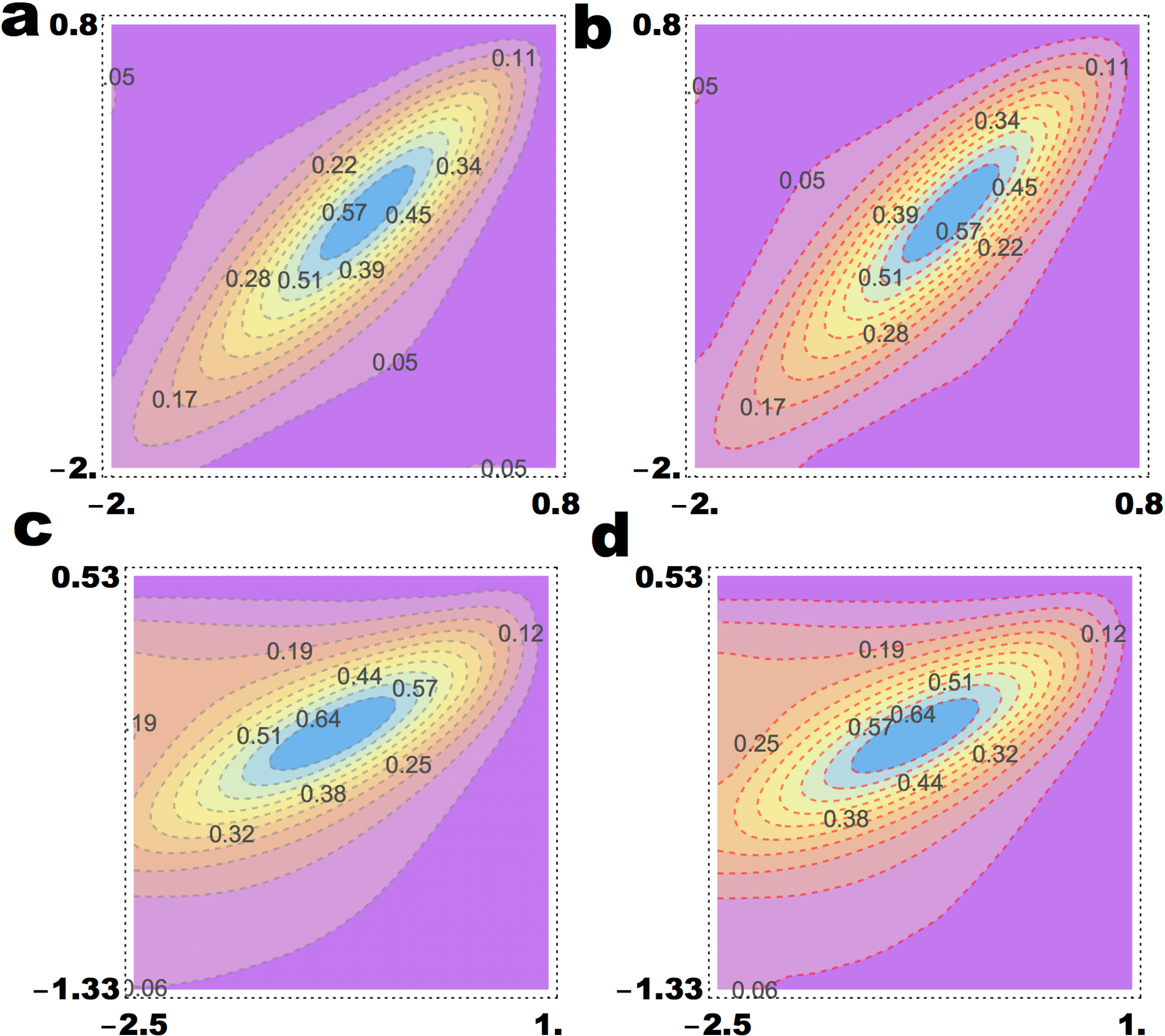}
		\caption{\label{fig4}  Joint membrane potential distributions of simulated data (smoothed 2D histograms of simultaneously recorded membrane potentials), compared to $P_0(x,y)$, Eq.~\ref{eq_2d_in_basis}, for $c=0.9$. The relative $L_1$ error is computed approximately as $0.02$. Parameters used for (a) and (b) are the same as Fig~\ref{fig2}. In (c) and (d) we present an asymmetric $\sigma_1$ and $\sigma_2$ pair. Dimensionless voltage boundaries are $x_r=-2.5$, $x_t=1.0$, $y_r=-1.33$, $y_t=0.53$.  The figures in the left column (gray dashed contours) are numerical renderings of our theory (solution of Eq.~\ref{eq_S_matrix}), whereas the figures in the right column (red dashed contours) are simulations of Eq.~\ref{eq_langevin_2D}. }
			\end{figure}
	
\section{Relation to Perturbative Approaches.}The perturbative solution for small 
$c$ is $S=S_0+c S_1+c^2 S_2+\ldots$. Inserting this into Eq.~\ref{eq_S_matrix} we obtain
\begin{multline}
\tilde{c} X (S_0+cS_1+c^{2}S_2+...) Y + \Lambda_1 (S_0+cS_1+\ldots)\\
+ (S_0+cS_1+\ldots) \Lambda_2 = -\tilde{c} X_0 Y_0 .
\end{multline}
We find that $S_0=0$ for $c=0$, since $\Lambda_{1k} S_{0,kl}+\Lambda_{2l} S_{0,kl} = 0$
has no nonzero solution with $\lambda_{1k} \neq -\lambda_{2k}$, except $\lambda_{1k} = 0 = \lambda_{2k}$ in which case we have set the coefficient of $f_0 g_0$ to $1$. The $O(c)$ equation for $S_1$ is
\begin{equation}
\Lambda_1 S_1 + S_1 \Lambda_2 = -\frac{1}{\sqrt{\tau_1\tau_2}}X_0 \otimes Y_0
\end{equation}
and using the definition $\psi_{kl} \equiv \frac{\sqrt{\tau_1\tau_2}}{\lambda_{1k}\tau_2+\lambda_{2l}\tau_1}$ the solution is 
\begin{equation}
S_{1,kl} = -\psi_{kl} X_{0,k} Y_{0,l}.
\end{equation}
The recursion relation for terms of order $O(c^n)$ is
$S_{n,kl} = -\psi_{kl}\sum_{ij} X_{ki} Y_{lj} S_{n-1,ij}$
with which one can expand the full perturbative series. Instead, for the non-perturbative regime, $S$ is obtained by solving a tensor equation
 \begin{eqnarray}
 &\sum_{kl} M_{ijkl}S_{kl} = F_{ij}\\
  & M_{ijkl} = \tilde{c}X_{ik} Y_{jl} + (\Lambda_{1i} + \Lambda_{2j}) \delta_{ik} \delta_{jl}\\
  & F_{ij} = -\tilde{c}X_{0,i} Y_{0,j}
  \end{eqnarray} 
which can be obtained by flattening indices and using conventional linear algebra techniques (Fig.~\ref{fig3}a).

\begin{figure}
	\includegraphics[width=8.6cm]{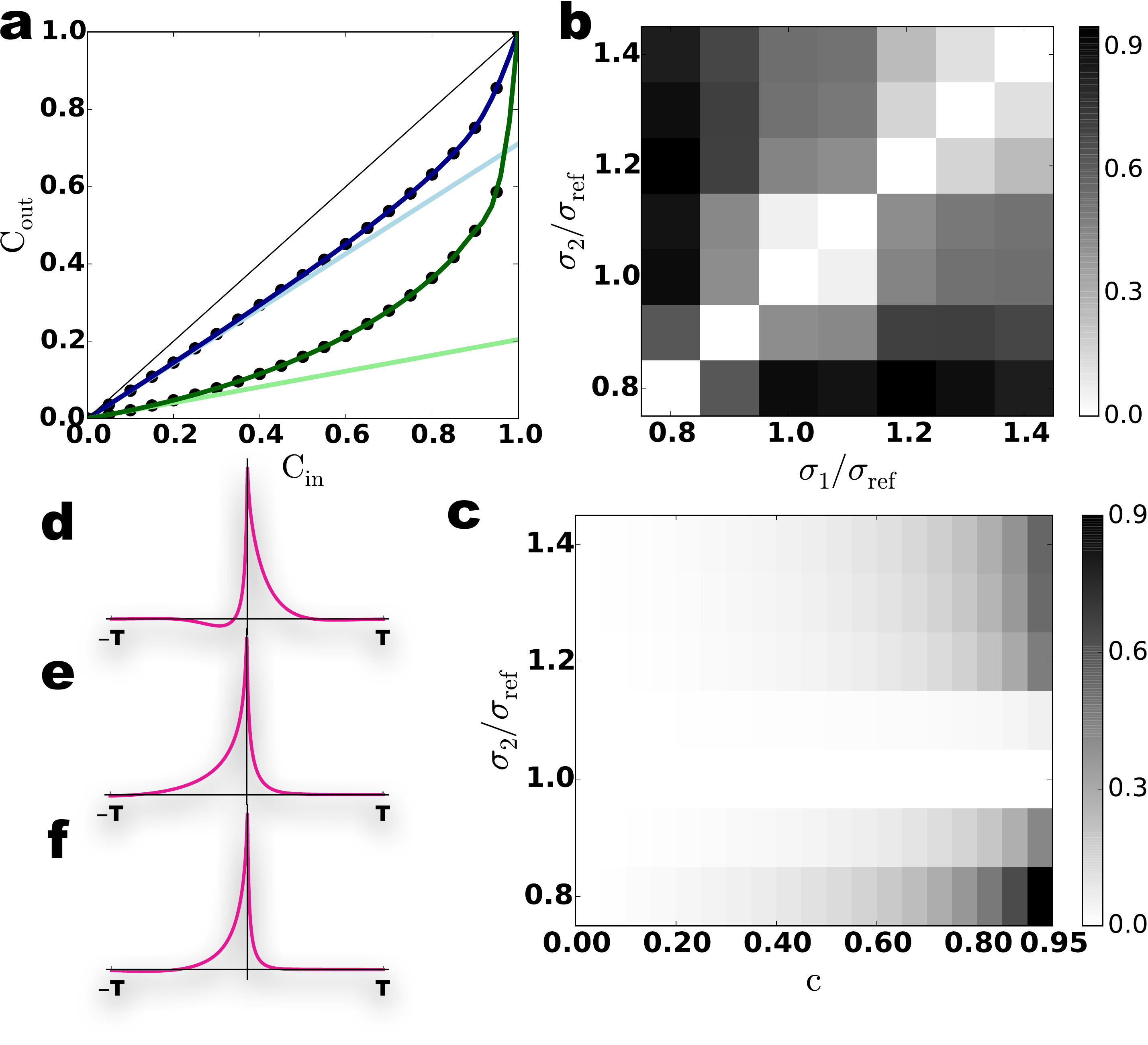}
	\caption{ Heterogeneous parameters lead to nonsymmetric cross-correlation functions. \label{fig5}(a)~Non-perturbative correlation transfer functions  $C_{out}(C_{in})$ in Eq.~\ref{eq_corr_coeff} for symmetric parameters and for high and low firing rates, respectively (blue: $r_b= \frac{0.231}{\tau_1}$, $\CV^2 =  0.5$; green: $r_g=\frac{0.017}{\tau_1}$, $\CV^2 = 0.98$). Slopes of light blue and light green lines (corresponding to $\frac{dC_{out}}{dC_{in}}$ at $C_{in}=0$), are computed using perturbation theory as in \cite{DelaRocha2007}. 
		(b)~Asymmetry of the cross-covariance function $\mathcal{A}=\int_{0}^{\infty}d\tau |C_{21}(\tau)-C_{21}(-\tau)|$ for two different input variances $\sigma_1$ vs.~$\sigma_2$, for $c=0.9$. (c)~$\mathcal{A}$ for changing input variance $\sigma_2$, fixed $\sigma_1=\sigma_{ref}$ and different values of $c$ between $0$ and $0.95$. Examples of asymmetric cross covariance functions (time rescaled with $\tau_1$ as in Fig.~\ref{fig2}d, $c=0.9$, time window $T=2.$) for heterogenous parameters in Eq.~\ref{eq_fokker_planck_2D} : (d)~asymmetric mean input $\mu_a$,   (e)~asymmetric membrane time constant $\tau_a$,  (f)~asymmetric input variance $\sigma_a^2$. }
\end{figure}

\section{Asymmetric Correlations.}Neurons in biological networks
have widely distributed parameters, and this heterogeneity may also
influence information processing \cite{Yim2013, Yim2012,Padmanabhan2010}. Moreover,
robust asymmetries in spike correlations could lead to asymmetric
synaptic efficacies when integrated via linear spike timing dependent
plasticity \cite{Morrison2008, Babadi2013}. Our approach reveals a
temporal asymmetry in covariance functions, Eq.~\ref{eq_cov_func}
related to a heterogeneity of intrinsic neuron parameters and input
parameters (Fig.~\ref{fig5}b). Such temporal asymmetry is more
pronounced for large values of $c$, especially in the non-perturbative
regime that we address in this work (Fig.~\ref{fig5}b--f.) (See Appendix for parameters.)

\section{Discussion and Conclusions.}We developed a novel theory of
correlation functions for two LIF model neurons driven by shared
input. Our approach can deal with the full range of input correlations
$0\leq c<1$, and the expansion converges fast (Fig.~\ref{fig3}b,d).
Also, our method is widely generalizable \cite{Rosenbaum2012}. Low
output firing rates generally require a non-perturbative treatment,
while the approximation derived from linear response theory
\cite{DelaRocha2007} is reasonably precise if firing rates are high
(Fig.~\ref{fig5}a). We considered firing rates between $1$ and
$25\,\mathrm{Hz}$, and values for $\CV^2$ between $0.5$ and $1$,
consistent with what is reported in neocortical neurons \textit{in
  vivo}. Strong correlations of membrane potentials were observed in
nearby neurons of cortical networks \cite{Lampl1999, Okun2008,Poulet2008}, 
compatible with the high degree of shared input
suggested from neuroanatomical studies. In the strongly correlated
regime the correlation transfer function is non-linear
\cite{DelaRocha2007, Schultze-Kraft2013} and the dynamics is quite
sensitive to heterogeneities of the input and of the model parameters
\cite{Yim2013, Yim2012,Padmanabhan2010}. Recent experiments demonstrated that asymmetric
correlation functions arise in neocortical neurons as well
\cite{Yim2013, Yim2012,Padmanabhan2010}. Correlation asymmetries could make an important
contribution to structure formation in networks through Hebbian
learning on short time scales in the range of the membrane time
constant of neurons \cite{Morrison2008,Babadi2013}.
\\

\section*{Acknowledgements.} We thank Man-Yi Yim for discussions. 
Funding by the BMBF (grant BFNT
01GQ0830) and the DFG (grant EXC 1086) is gratefully acknowledged.


	\section*{Appendix   A: Eigenvalue spectrum of 1D operators}
	
		This section clarifies results of the main text  and includes detailed step by step computations.  We use shorthand notations for eigenfunctions, $f_i(x)\equiv f_{\lambda_i}(x)$ interchangably. We repeat some equations of the main text in order to put detailed computations in context.
	
	Two independent solutions of the following Sturm-Liouville problem
	\begin{equation}
	\label{sup_eq_sturm_leuville}
	\mathcal{L}_1\phi=\frac{\partial (x\phi)}{\partial x}+\frac{1}{2}\frac{\partial^2 \phi}{\partial x^2} =\lambda \phi
	\end{equation}
	are given in \cite{Abramowitz:1974:HMF:1098650} as
	\begin{equation}
	\phi_1(x,\lambda)=\: _1F_1\big{(}\frac{1-\lambda}{2},1/2,-x^2\big{)} 
	\end{equation}
	\begin{multline}
	\phi_2(x,\lambda)=\frac{\Gamma(\frac{\lambda}{2})}{\Gamma(\frac{\lambda+1}{2})} \: _1F_1\big{(}\frac{1-\lambda}{2},\frac{1}{2},-x^2\big{)} +\\
	2x \: _1F_1\big{(}1-\frac{\lambda}{2},\frac{3}{2},-x^2\big{)} 
	\end{multline}
	
	where $_1F_1(a,b,z)$ is the Confluent Hypergeometric Function of the first kind \cite{Abramowitz:1974:HMF:1098650}. We note that the fraction $\frac{\Gamma(\frac{\lambda}{2})}{\Gamma(\frac{\lambda+1}{2})}$ is regularized, as the reciprocal of gamma functions can be analytically continued to zero at its poles \cite{Abramowitz:1974:HMF:1098650}. We note that there is another basis known to be numerically stable, given in terms of Parabolic Cylinder Functions \cite{Lindner2001, Abramowitz:1974:HMF:1098650} 
	\begin{align}
	\psi_1(x,\lambda)&=\: e^{-\frac{x^2}{2}}D_{-\lambda}(x/\sqrt{2})\\
	\psi_2(x,\lambda)&=e^{-\frac{x^2}{2}}D_{\lambda-1}( i x/\sqrt{2}).
	\end{align}
	 It doesn't matter which basis is used to expand a function in the eigenspace of $\mathcal{L}_1$.  Eigenfunctions are unique up to some normalization condition which we select to be $R(\lambda)=-\frac{1}{2}\partial_x{f_\lambda(x_t)}=1$. The eigenvalue spectrum of Eq.~\ref{sup_eq_sturm_leuville} is discrete and can be found by satisfying the boundary conditions
	\begin{equation}
	\begin{split}
	\label{sup_sup_eq_bc_1}
	f_\lambda(x_{t})=0=\lim_{x\rightarrow -\infty} f_\lambda(x) \\
	f_\lambda(x_r-\epsilon)\stackrel{\epsilon \rightarrow 0}{= }  f_\lambda(x_r+\epsilon)\\
	\partial_x f_\lambda(x_{t})\stackrel{\epsilon \rightarrow 0}{= } \partial_x f_\lambda(x_r-\epsilon)-\partial_x f_i(x_r+\epsilon).
	\end{split}
	\end{equation}
		\begin{widetext}
	A general family of solutions with the property $\lim_{x\rightarrow -\infty} f_\lambda(x)=0$ is given as
	\[ f_\lambda(x)=\begin{cases} 
	a(\lambda) \phi_1(\lambda,x) +b(\lambda) \phi_2(\lambda,x) &x_r \leq x < x_{t} \\
	d(\lambda) \phi_2(\lambda,x)& x_r \geq x
	\end{cases}.
	\]
	The boundary conditions Eq.\ref{eq_bc_1} require 
	\begin{eqnarray*}
		a(\lambda) \phi_1(\lambda,x_{t}) +b(\lambda) \phi_2(\lambda,x_{t})=0\\
		a(\lambda) \phi_1(\lambda,x_{r}) +b(\lambda) \phi_2(\lambda,x_{r})-d(\lambda) \phi_2(\lambda,x_r)=0\\
		a(\lambda) (\phi_1'(\lambda,x_{r})-\phi_1'(\lambda,x_{t})) +b(\lambda) (\phi_2'(\lambda,x_{r})- \phi_2'(\lambda,x_{t}))-d(\lambda) \phi_2'(\lambda,x_r)=0
	\end{eqnarray*}
	 and in order to have non-zero solutions the determinant of the coefficient matrix must satisfy
	\begin{equation}
	\begin{vmatrix}
	\phi_1(\lambda,x_{t})  & \phi_2(\lambda,x_{t})& 0\\ 
	\phi_1(\lambda,x_{r})  & \phi_2(\lambda,x_{r}) &  -\phi_2(\lambda,x_r) \\
	(\phi_1'(\lambda,x_{r})-\phi_1'(\lambda,x_{t})) \: & (\phi_2'(\lambda,x_{r})- \phi_2'(\lambda,x_{t})) \: & -\phi_2'(\lambda,x_r)
	\end{vmatrix}
	=0.
	\end{equation}

	The eigenvalues $\{\lambda_i\}$ are countably many isolated points given as solutions of 
	\begin{equation}
	\label{sup_eq_spectral}
	\phi_2(\lambda, x_{t})\:  /Wr(x_{t})-\phi_2(\lambda, x_r)\: /Wr(x_{r})=0
	\end{equation}
	where we have the Wronskian $Wr(x)= \phi_1'(x) \phi_2(x)-\phi_1(x) \phi'_2(x)=2e^{-x^2} $. The spectrum is the
	same as given in \cite{Brunel1999}. In order to find $a$ and $b$, we need to fix $d(\lambda)$
	\begin{equation*}
	\begin{split}
	a(\lambda)=
	\frac{\phi_2(\lambda,x_r)e^{x_r^2}\: d(\lambda)}{e^{x_r^2}\phi_1(\lambda, x_r)-e^{x_{t}^2}\phi_1(\lambda,x_{t})}
	\end{split}
	\end{equation*}
	\begin{equation*}
	\begin{split}
	b(\lambda)=\frac{-\phi_1(\lambda,x_{t})e^{x_{t}^2}\: d(\lambda)}{e^{x_r^2}\phi_1(\lambda, x_r)-e^{x_{t}^2}\phi_1(\lambda,x_{t})}
	\end{split}.
	\end{equation*}
	We can find the exit rate at threshold $R(\lambda)$ as 

	\begin{align*}
	R(\lambda)=-\frac{1}{2}\partial_x{f(\lambda,x_t)}&=-\frac{1}{2}(a(\lambda)\phi'_1(\lambda, x_{t})+b(\lambda)\phi'_2(\lambda, x_{t}))\\
	&=-\frac{1}{2}d(\lambda)\frac{\phi_2(\lambda,x_{r})\phi'_1(\lambda,x_{t})e^{x_r^2}-\phi'_2(\lambda,x_{t})\phi_1(\lambda,x_{t})e^{x_{t}^2}}{e^{x_r^2}\phi_1(\lambda, x_r)-e^{x_{t}^2}\phi_1(\lambda,x_{t})}\\
	&=-\frac{1}{2}d(\lambda)\frac{\phi_2(\lambda,x_{t})\phi'_1(\lambda,x_{t})e^{x_t^2}-\phi'_2(\lambda,x_{t})\phi_1(\lambda,x_{t})e^{x_{t}^2}}{e^{x_r^2}\phi_1(\lambda, x_r)-e^{x_{t}^2}\phi_1(\lambda,x_{t})}\\
	&=-\frac{1}{2}d(\lambda)\frac{e^{x_t^2}(\phi_2(\lambda,x_{t})\phi'_1(\lambda,x_{t})-\phi'_2(\lambda,x_{t})\phi_1(\lambda,x_{t}))}{e^{x_r^2}\phi_1(\lambda, x_r)-e^{x_{t}^2}\phi_1(\lambda,x_{t})}\\
	&=-\frac{1}{2}d(\lambda)\frac{-e^{x_t^2} \: 2e^{-x_t^2}}{e^{x_r^2}\phi_1(\lambda, x_r)-e^{x_{t}^2}\phi_1(\lambda,x_{t})}\\
	&=\frac{d(\lambda)}{e^{x_r^2}\phi_1(\lambda, x_r)-e^{x_{t}^2}\phi_1(\lambda,x_{t})}
	\end{align*}
	
	where we select 
	\begin{equation} 
	d(\lambda)=e^{x_r^2}\phi_1(\lambda, x_r)-e^{x_{t}^2}\phi_1(\lambda,x_{t})
	\end{equation}
	in order to have $R(\lambda)=1$. As a result we obtain 
	\begin{equation}
	a(\lambda)=\phi_2(\lambda,x_r)e^{x_r^2}
	\qquad\text{and}\qquad
	b(\lambda)=-\phi_1(\lambda,x_{t})e^{x_{t}^2}.
	\end{equation}
We note that there is a numerical method which generalizes the procedure above to neuron models with no known explicit solutions 
	\cite{Richardson2008, Ostojic2011}. 
\end{widetext}
	\section*{Appendix   B: Dual eigenspace}
	
	In this section we explain non-orthogonal projections to a non-adjoint operator eigenspace.  Again we use shorthand notations for eigenfunctions, $f_i(x)\equiv f_{\lambda_i}(x)$ interchangably. The solution to the Sturm-Liouville equation, $f_\lambda(x)$, satisfying
	\begin{equation}
	\mathcal{L}_1f=\frac{\partial (xf)}{\partial x}+\frac{1}{2}\frac{\partial^2 f}{\partial x^2} =\lambda f
	\end{equation}
	are given above. As $\mathcal{L}_1$ is not an adjoint operator (because of reset boundary conditions in Eq.~\ref{eq_bc_1} ), in order to build a bi-orthogonal basis, we need to find the dual equation $\mathcal{L}^{\dagger}f=\lambda f$ \cite{risken-fpe-1996}, which satisfies
	\begin{align}
	\langle \tilde{f}_j	\mathcal{L}_1^{\dagger},f_i\rangle -\langle \tilde{f}_j,\mathcal{L}_1 f_i\rangle=(\lambda_j-\lambda_i) \langle f_i,\tilde{f}_j \rangle
	\end{align}
	where $\langle .,.\rangle $ is an inner product in Hilbert space which is given in \cite{risken-fpe-1996} explicitly as
	\begin{align}
	\int	dx f_i\mathcal{L}_1^{\dagger} \tilde{f}_j-\int dx \tilde{f}_j\mathcal{L}_1 f_i=(\lambda_j-\lambda_i) \int dx f_i\tilde{f}_j .
	\end{align}
	Here the LHS is the surface term which can be simplified by integration by parts as
	\begin{align}
	(\lambda_j-\lambda_i) \int dx f_i\tilde{f}_j =- [\tilde{f}_j J_j]_{-\infty}^{x_r}-[\tilde{f}_j J_j]_{x_r}^{x_{t}}- [\partial_x \tilde{f}_j f_j]_{-\infty}^{x_{t}}
	\end{align}          
	where we defined the current $ J_i\equiv -xf_i-\frac{1}{2}\partial_xf_i$ and $[f(x)]^{a}_{b}\equiv f(a)-f(b)$.  Dual boundary conditions that satisfy zero surface term are then  
	\begin{eqnarray}
	\tilde{f}_i(x_r)=\tilde{f}_i(x_{t}). 
	\end{eqnarray}
	This guarantees that $\langle f_i,\tilde{f}_j \rangle =\delta_{ij}$ with appropriate choice of constants. The corresponding dual equation is 
	\begin{eqnarray}	
	\label{sup_eq_dual_equation}
	\mathcal{L}_1^{\dagger}\tilde{f}=-x\frac{\partial \tilde{f}}{\partial x}+\frac{1}{2}\frac{\partial^2 \tilde{f}}{\partial x^2} =\lambda \tilde{f}\: \text{.}
	\end{eqnarray}
	The transformation $\tilde{f}(x)=e^{x^2}h(x)$ with following relations
	\begin{align*}
	\tilde{f}'&=\Big{(}2xh+h'\Big{)} e^{x^2}\\
	\tilde{f}''&=\Big{(}(2+4^2x^2)h+4xh'+h''\Big{)} e^{x^2}\\
	-x\tilde{f}'&=\Big{(}-2x^2h-xh'\Big{)} e^{x^2}
	\end{align*}
	with $\tilde{ f}_i$ satisfying Eq.~\ref{sup_eq_dual_equation} for an eigenvalue $\lambda_i$ . It can be shown after insertion of equations above in Eq.~\ref{sup_eq_dual_equation} that
	\begin{eqnarray}	
	\mathcal{L}_1h=\frac{\partial (xh)}{\partial x}+\frac{1}{2}\frac{\partial^2 h}{\partial x^2}=\lambda h
	\end{eqnarray}
	holds. The dual eigenfunctions are found to be
	\begin{equation}
	\tilde{f}_i(x)=e^{x^2}(\tilde{a}(\lambda_i) \phi_1(\lambda_i,x) +\tilde{b}(\lambda_i) \phi_2(\lambda_i,x)).
	\end{equation}
	The boundary conditions require that continuous and differentiable solutions satisfy  
	\begin{multline}
	e^{x_r^2}(\tilde{a}(\lambda_i) \phi_1(\lambda_i,x_r) +\tilde{b}(\lambda_i) \phi_2(\lambda_i,x_r)) = \\
	e^{x_{t}^2}(\tilde{a}(\lambda_i) \phi_1(\lambda_i,x_{t}) +\tilde{b}(\lambda_i) \phi_2(\lambda_i,x_{t})).
	\end{multline}
	This implies that $\tilde{a}=0$ because of the spectral equation Eq.~\ref{sup_eq_spectral}, and as a nonzero Wronskian implies the
	independence of two solutions.
	Finally, we select $\tilde{b}(\lambda_i)$ such that $\langle \tilde{f_i}, f_j\rangle=\delta_{ij}$,
	\begin{equation}
	\tilde{f}_i(x)=\frac{e^{x^2}\phi_2(\lambda_i,x)}{\langle e^{x^2}\phi_2(\lambda_i,x),f_i(x)\rangle} \: \: .
	\end{equation}
	
\section*{Appendix   C: Details of the series expansion}
		
		This section repeats results of the main text  and includes detailed step by step computations.  We use again a shorthand notation for eigenfunctions, $f_i(x)\equiv f_{\lambda_i}(x)$. We repeat equations of the main text in order to put detailed computations in context.
		
		In order to investigate regularized reset boundary conditions, we write derivatives of eigenfunctions in the form
		\begin{subequations}
			\label{sup_eq_regularized_bc}
			\begin{eqnarray}
			\partial_xf_0(x) = 2r_1\tau_1[\kappa(x)+\sum_{k=1} X^{(1)}_{0k}    f_k(x)   ]+ A    f_0(x)    \qquad \\ 
			\partial_yg_0(y) = 2r_2\tau_2[\kappa(y)+\sum_{l=1} Y^{(1)}_{0l}  g_l(y)  ]+ B    g_0(y)     \qquad  \\
			\partial_xf_i(x) = R_{1i}\kappa(x)+\sum_{k=1}  X^{(1)}_{ik}    f_k(x)   + \frac{A}{2r_1\tau_1}    f_0 (x)    \qquad \\ 
			\partial_yg_j(y) = R_{2j}\kappa(y)+\sum_{l=1}  Y^{(1)}_{jl}  g_l(y)+ \frac{B}{2r_2\tau_2}    g_0(y)  \qquad   
			\end{eqnarray}
		\end{subequations}
		where $X^{(1)}$ are generalized Fourier coefficients of a continuous function $\partial \bar{f}_i =\partial_xf_i -R_{1i}\kappa(x)$ and similarly for $Y^{(1)}$.
		The constants $R$ defined above are chosen as
		\begin{eqnarray}
		R_{1i}=-\frac{1}{2}\partial_xf_i \big{|}_{x_{t}}=1\\
		R_{2j}=-\frac{1}{2}\partial_yg_j \big{|}_{y_{t}}=1
		\end{eqnarray}
		
		The box function $\kappa$ is defined as
		\begin{eqnarray}
		\kappa(x)=\Theta(x-x_r)-\Theta(x-x_{t}) 
		\end{eqnarray}
		with Heaviside functions
		
		\[ \Theta(x)=\begin{cases} 
		0 & x \leq 0 \\
		1 & x > 0 
		\end{cases}.
		\]
		It should be pointed out that one encounters an analog of the ``Gibbs phenomenon'' for generalized Fourier series for 
		our case of a non-selfadjoint series expansion \cite{Adcock2012}. This partially limits the convergence properties of our theory.

		One can easily show via direct integration and using boundary conditions 
		\begin{eqnarray*}
			\int^{x_t}_{-\infty}dx \: \tilde{f}_0\partial_x f_i(x)  =0  \\ 
			\int^{y_t}_{-\infty}dx \: \tilde{g}_0\partial_y g_i(y)  =0 .
		\end{eqnarray*}
		
		This implies that the projections $ \tilde{f}_0 \tilde{g}_0$ ,$ \tilde{f}_0 \tilde{g}_l$, $ \tilde{f}_k \tilde{g}_0 $ are identically zero. 
		Hence, the constants $A$ and $B$ are found as 
		\begin{eqnarray}
		A=-\int^{x_{t}}_{x_r} \tilde{f}_0=x_{r}-{x_t}\\
		B=-\int^{y_{t}}_{y_r} \tilde{g}_0=y_{r}-{y_t}
		\end{eqnarray}
		The solution as a series expansion in the basis above is
		\begin{equation}
		\label{sup_eq_series_expansion_2d}
		P_0(x,y)=f_0(x)g_0(y)+F(x) SG(y)
		\end{equation}
		where we define  $F(x) SG(y)=\sum_{ij} S_{ij}f_i(x)g_j(y)$. The first column and first row of the expansion coefficients are zero except the coefficient of $f_0g_0$, leaving only the
		matrix $S$ with $S_{ij} \in \mathbb{C}$ as unknown. This expansion satisfies the constraints for marginal distributions
		
		\begin{multline}
		\int_{-\infty}^{y_{t}} dyP_0(x,y)=\\
		f_0(x) \int_{-\infty}^{y_{t}} dy\: g_0(y)+\sum_{ij} S_{ij}f_i(x)\int_{-\infty}^{y_{t}} dy \: g_j(y)=f_0(x) 
		\end{multline}
	
		as $\int_{-\infty}^{y_{t}} dy\: g_0(y)=1$ and $\int_{-\infty}^{y_{t}} dy \: g_j(y)=0$. The probability distribution $f_0(x)$ is given by
		\begin{eqnarray}
		f_0(x)=2r_1\tau_1e^{-x^2}\int_{x}^{x_{t}} du \: \Theta(u-x_r)e^{u^2}.
		\end{eqnarray}
		A constraint for $g_0(y)$ is given analogously. Again, $\Theta(x)$ is the Heaviside function.
		Using $\int_{-\infty}^{x_{t}}dx\: f_0(x)=1$ and changing variables, steady state rates are as in Eq~\ref{eq_steady_state_rate}
		We obtain the same expression for $r_2$ with the appropriate parameters. Using Eq.\ref{eq_1d_eigenspace_1},  Eq.~\ref{eq_2d_stationary}  is given in terms of eigenfunctions as 
		\begin{equation}
		\label{sup_eq_2d_in_basis}
		\begin{split}
		F(x) \Lambda_1 S G(y)+F(x) S \Lambda_2  G(y)+\tilde{c} \partial_x F(x) S \partial_y G(y)=\\
		-\tilde{c}\partial_x f_0(x) \partial_y g_0(y)
		\end{split}
		\end{equation}
		with $\Lambda_a=\frac{\lambda_{ai}}{\tau_a}\delta_{ij}$ and $\tilde{c}=\frac{c}{\sqrt{\tau_1\tau_2}}$. In order to solve Eq.~\ref{sup_eq_2d_in_basis} we express the action of derivative operators on the eigenbasis as 
		\begin{align}
		X_{ij} &= \int_{-\infty}^{x_{t}} \tilde{f}_i(x) \partial_xf_j(x)\,dx  \\
		Y_{ij} &= \int_{-\infty}^{y_{t}} \tilde{f}_i(y) \partial_yf_j(y)\,dy \\
		X_{i0} &= \int_{-\infty}^{x_{t}} \tilde{f}_0(x) \partial_xf_j(x)\,dx  \\
		Y_{i0} &= \int_{-\infty}^{y_{t}} \tilde{f}_0(y) \partial_yf_j(y)\,dy .
		\end{align}
		The final equation in matrix form is then
		\begin{equation}
		\label{sup_eq_S_matrix}
		\Lambda_1 S +  S \Lambda_2  +\tilde{c} X^{T}S Y=-\: \tilde{c}\:X_0 \otimes Y_0.
		\end{equation}
		Here we should note that we solve an equation assuming stationarity in a discrete sub-space.  
		This is only an approximation of the unique full solution of Eq.~\ref{eq_fokker_planck_2D}.
		In this way,  we can obtain an approximate solution (due to sub-space projections) with arbitrary precision.  
		The way we constructed this solution provides us with explicit spike train covariance functions. 
		The covariance function of two stationary spike trains represented as a sum of delta functions $\rho_1 = \sum_k \delta(t-t^1_k)$ and $\rho_2 = \sum_l \delta(t-t^2_l)$ is given as
		\begin{equation}
		\begin{split}
		C_{ij} (\tau)= \langle\rho_1(t+\tau) \rho_2(t) \rangle \\
		-\langle\rho_1(t+\tau)\rangle \langle \rho_2(t) \rangle
		\end{split}
		\end{equation}
		can be simplified in terms of the conditional rate $r_{i|j}(\tau)$ as 
		\begin{equation}
		\label{sup_eq_corr_rate}
		C_{ij} (\tau)= r_i(r_{i|j}(\tau)-r_j).
		\end{equation}
		For any given stationary joint membrane potential distribution $P_0(x,y)$, the distribution of the membrane potential conditional to a spike at $t_0=0$ is expressed as
		\begin{equation}
		P_{a|b}(x)=\mathrm{Pr}(x \, | \, \text{spike in $[t_0,t_0+dt)$}).
		\end{equation}
		The conditional probability of observing a spike in the sequel is then
		$P_{1|2}(x)=-\frac{1}{2r_2\tau_2}\partial_{y}P_0(x,y_{t})$    
		as $\int_{-\infty}^{x_{t}} dx \: \partial_{y}P_0(x,y_{t})=-2r_2\tau_2$ by construction. Solving the initial value problem 
		\begin{eqnarray}
		f(t_0,x)=-\frac{1}{2r_2\tau_2}\partial_{y}P_0(x,y_{t})\\
		\partial_t f=\mathcal{L}_1f
		\end{eqnarray}
		where $\mathcal{L}_1$ is the time evolution operator in Eq.~\ref{eq_1d_eigenspace_1}. Using Eq.~\label{eq_series_expansion_2d}, the explicit solution for $P_0(x,y)$, the instantaneous conditional distribution is found as  
		\begin{equation}
		\begin{split}
		P_{1|2}(x)=-\frac{1}{2r_2\tau_2}\partial_{y}P_0(x,y_{t})=\\
		f_0-\frac{1}{2r_2\tau_2}\sum_i f_i(x)\Big{(}\sum_j S_{ij}\Big{)},   
		\end{split}
		\end{equation}  
		because $\partial_{y}g_0(y_{t})=-2r_2\tau_2$ and $\partial_{y}g_i(y_{t})=-1$. Applying the time evolution operator
		\begin{align*}
		f(x,t) &=e^{\mathcal{L}_1t}[f_0-\frac{1}{2r_2\tau_2}\sum_i f_i(x)\Big{(}\sum_j S_{ij}\Big{)}   ]\\
		&=f_0-\frac{1}{2r_2\tau_2}\sum_i f_i(x)\Big{(}\sum_j S_{ij}\Big{)}   e^{\Lambda _{1i} t} 
		\end{align*}  
		the conditional rate becomes
		\begin{align*}
		r_{12}(t)&=-\frac{1}{2\tau_1}\partial_x f(x_t,t) \\
		&=r_1+\frac{1}{4r_2\tau_2\tau_1}\sum_i \Big{(}\sum_j S_{ij}\Big{)} e^{\Lambda _{1i} t} .
		\end{align*}  
		Using this in Eq. \ref{sup_eq_corr_rate} yields
		\begin{equation}
		C_{12} (\tau)= r_2(r_{1|2}(\tau)-r_1)=\frac{1}{4\tau_2\tau_1}\sum_i e^{\Lambda _{1i} t} \Big{(}\sum_j S_{ij}\Big{)} 
		\end{equation}
		The counterpart of this is computed in a similar way
		\begin{eqnarray}
		\label{sup_eq_c_21}
		C_{12}(\tau)=\frac{1}{4\tau_2\tau_1}\sum_j \Big{(}\sum_i S_{ij}\Big{)} e^{\Lambda _{2j}\tau} 
		\end{eqnarray}
		Finally, the integral of the covariance is then found as
		\begin{eqnarray}
		\int_{-\infty}^{\infty}C(\tau)=\sum_{ij}(S\Lambda_2^{-1}+\Lambda_1^{-1} S)_{ij}
		\end{eqnarray}
		by reordering the matrices and using Eq. \ref{sup_eq_S_matrix}
		\begin{align}
		\int_{-\infty}^{\infty}C(\tau)&=\sum_{ij}(\Lambda_1^{-1}\Lambda_1 S \Lambda_2^{-1}+\Lambda_1^{-1} S\Lambda_2 \Lambda_2^{-1}  )_{ij}\\
		&=\sum_{ij}(\Lambda_1^{-1}(-\tilde{c} XS Y-\tilde{c}X_0Y_0) \Lambda_2^{-1}  )_{ij}\\
		&=-\tilde{c}\sum_{ij}(\Lambda_1^{-1}(XS Y+X_0Y_0) \Lambda_2^{-1}  )_{ij}
		\end{align}

\section*{Appendix  D:  Comparison to linear response theory}
		
		The perturbative solution for small 
		$c$ is  given as a geometric series with matrix coefficients $S=S_0+c S_1+c^2 S_2+...$. Inserting this into Eq.~\ref{sup_eq_S_matrix} we obtain
		\begin{multline}
		\tilde{c} X (S_0+cS_1+c^{2}S_2+...) Y + \Lambda_1 (S_0+cS_1+...)+ \\
		(S_0+cS_1+…) \Lambda_2 = -\tilde{c} X_0 Y_0 .
		\end{multline}
		We find that $S_0=0$ for $c=0$, since $\Lambda_{1k} S_{0,kl}+\Lambda_{2l} S_{0,kl} = 0$
		has no nonzero solution with $\lambda_{1k} \neq -\lambda_{2k}$, except $\lambda_{1k} = 0 = \lambda_{2k}$ in which case we have set the coefficient of $f_0 g_0$ to $1$. The $O(c)$ equation for $S_1$ is
		\begin{equation}
		\Lambda_1 S_1 + S_1 \Lambda_2 = -\frac{1}{\sqrt{\tau_1\tau_2}}X_0 \otimes Y_0
		\end{equation}
		
		and using the definition $\psi_{kl} \equiv \frac{\sqrt{\tau_1\tau_2}}{\lambda_{1k}\tau_2+\lambda_{2l}\tau_1}$ the solution is 
		\begin{equation}
		S_{1,kl} = -\psi_{kl} X_{0,k} Y_{0,l}.
		\end{equation}
		The recursion relation for terms of order $O(c^n)$ is
		$S_{n,kl} = -\psi_{kl}\sum_{ij} X_{ki} Y_{lj} S_{n-1,ij}$
		with which one can expand the full perturbative series.

		The result of linear response theory for output spike train correlations is given in \cite{Shea-Brown2008} as
		\begin{widetext}
		\begin{multline}
		C_{\mathrm{out},\mathrm{pert}}^{(1)}=\frac{c\sigma_1 \sigma_2 \frac{dr_1}{d\mu}\frac{dr_2}{d\mu}}{\CV_1 \CV_2\sqrt{r_1r_2} } =\frac{c \: r_1^{3/2} r_2^{3/2} [e^{x_t^2} \erf(x_t)-e^{x_r^2} \erf(x_r)][e^{y_t^2} \erf(y_t)-e^{y_r^2} \erf(y_r)]}{\CV_1 \CV_2}
		\end{multline}
		\end{widetext}
		where $\erf(x)$ is the error function
		\cite{Abramowitz:1974:HMF:1098650}. We used the following formula for the $CV^2=\frac{\sigma^2_{ISI}}{\mu^2_{ISI}}$,
		\begin{equation}
		CV^{2}=2\pi \nu^2 \int_{y_{res}}^{y_{th}}dx e^{x^2}\int^{y}_{-\infty}dy[1+\erf(x)]^2
		\end{equation}
		given in \cite{Brunel2000}. We compare this to our result
		(shown in Fig.~\ref{fig5}a)
		\begin{multline}
		C^{(1)}_\mathrm{out}\approx -\frac{c}{4\sqrt{\tau_1\tau_2}}\sum_{ij}\Big{[}\Lambda_1 ^{-1}X_0Y_0\Lambda_2^{-1} \Big{]}_{ij}\\
		+O(c^2)
		\end{multline}
		and find a perfect match. Moreover, $C_\mathrm{out}$ with quadratic corrections can be easily calculated
		\begin{multline}
		C^{(2)}_\mathrm{out}\approx -\frac{c}{4\sqrt{\tau_1\tau_2}}\sum_{ij}\Big{[}\Lambda_1 ^{-1}(c X X_0 \psi Y_0 Y+X_0Y_0)\Lambda_2^{-1} \Big{]}_{ij}\\
		+O(c^3).
		\end{multline}
		where $\psi_{kl} \equiv \frac{\sqrt{\tau_1\tau_2}}{\lambda_{1k}\tau_2+\lambda_{2l}\tau_1}$.

		\section*{Appendix E: Numerical analysis and parameters}
		\subsection{Numerical evaluation of correlations }
		
		We compute spike train correlations via average conditional histograms. (We use \texttt{numpy.histogram()} to obtain the probability of $P(t^{a}_i-t^{b}_j)$ using a triangular envelope around zero lag, as weight function.)
		One can express this as an integral over 
		two variables $\tau=t_1-t_2$ and $s=t_1+t_2$ with bin size $\Delta$
		\begin{equation}
		\begin{split}
		C(\tau)=\frac{1}{\Delta}	\int_{\tau}^{\tau+\Delta} \frac{d\tau'}{u(\tau')-l(\tau')} 
		\int_{l(\tau')}^{u(\tau')} ds' \\
		\sum_{i,j} \delta(\tau'-\tau_i)\delta(s'-s_j)
		\end{split}
		\end{equation}
		where we have
		\[ u(\tau)=\begin{cases} 
		T/\sqrt{2}-\tau & \tau<0\\
		T/\sqrt{2}+\tau & \tau>0
		\end{cases}
		\]
		\[ l(\tau)=\begin{cases} 
		T/\sqrt{2}+\tau & \tau<0\\
		T/\sqrt{2}-\tau & \tau>0
		\end{cases}
		\]
		with observation window $T$. 
		
		\subsection{Solution of stochastic differential equations}
		
		We used Euler-Maruyama scheme to integrate stochastic differential equations, like the Ornstein-Uhlenbeck Process
		\begin{equation}
		\tau \dot{V}=-V+\mu+ \sigma\sqrt{\tau} [\sqrt{1-c}\:\xi\pm\sqrt{c}\:\xi_c].§
		\end{equation}
		The discrete time approximation with $t_0<t_1<t_2...<t_n<T$ is then 
		\begin{equation}
		V_{i+1}=(1-\frac{dt}{\tau})V_i+\frac{dt}{\tau}\mu+\sqrt{\frac{dt}{\tau}}[\sqrt{1-c}\: n_i \pm\sqrt{c}\:n_{c,i}]
		\end{equation}
		where $n_i$ and $n_{ci}$ are normally distributed random numbers $\sim\mathcal{N}(0,1)$. 
		
		\subsection{Voltage data and smoothing}
		We simulated the stochastic differential equation in \texttt{Python}. We recorded simulated data for several trials and binned 2D data with the function \texttt{numpy.histogram()}. We averaged the histogram for $N_\mathrm{trial}$ trials. We smoothed the histogram data 
		with a 2D boxcar kernel averaging over $m \times n$ bins. Parameters used are given in Tab.~\ref{sup_Tab:01}.
		
		\begin{table}[h!]
			\textbf{\refstepcounter{table}\label{sup_Tab:01} Table \arabic{table}.}{ Parameters for Fig.~\ref{fig2} and Fig.~\ref{fig3} }
			{\begin{tabular}{llr}
					\hline		
					\multicolumn{2}{c}{\textbf{ Model parameters}} \\
					\hline
					Symbol    & Description & Value  \\
					\hline
					$x_{t}$, $y_{t}$      & voltage threshold     & $0.8$      \\
					$x_{r}$, $y_{r}$     & voltage reset    & $-2.$   \\
					$\tau_m$ & membrane time constant      & $1.$        \\
					\hline
					\multicolumn{2}{c}{\textbf{ Simulation parameters}} \\
					\hline
					$dt$      & time bin     & $0.005$      \\
					$t_\mathrm{total}$    & total time     & $2000$   \\
					$N_\mathrm{trials}$ &  number of independent trials       & $20$       \\
					\hline
					\multicolumn{2}{c}{\textbf{ Data analysis parameters}} \\
					\hline
					$N_\mathrm{xbins}$      & number of bins in $x$ direction     & $300$     \\
					$N_\mathrm{ybins}$      & number of bins in $y$ direction  & $300$\\
					$[x_-,x_t]$ &  data recording range      & $[-3,0.8]$       \\
					&  2D boxcar smoothing range  & $10\times 10$ bins    \\
					\hline
					\multicolumn{2}{c}{\textbf{ Statistics of output spike trains }} \\
					\hline
					$r_{1}$,$r_{2}$      & spikes per $\tau_m$   & $0.231$     \\
					$\CV^2_1$, $\CV^2_2$      &   squared coefficient of variation & $0.5$ \\
					\hline
					\multicolumn{2}{c}{\textbf{ Numerical analysis of correlations }} \\
					\hline		
					$T_\mathrm{observe}$      & observation time interval & $[-2.,2.]$    \\
					$N_\mathrm{bins}$    &   number of bins & $\sim 450$
				\end{tabular}}{}
			\end{table}
	
			\begin{table}[h!]
				\textbf{\refstepcounter{table}\label{sup_Tab:02} Table \arabic{table}.}{ Parameters for Fig.~\ref{fig5}a}	
				
				{\begin{tabular}{llr}			
						\hline
						\multicolumn{2}{c}{\textbf{ Numerical analysis data }} \\
						\hline
						Symbol    & Description & Value  \\			
						\hline
						$C_{in}$ & range of input correlation data points  & $[0,0.95]$\\
						$\Delta C_{in}$ & step of input correlation data points  & $0.05$\\
						\hline		
						\multicolumn{2}{c}{\textbf{ Model 1 (dark blue) parameters}} \\
						\hline		
						$x_{t}$, $y_{t}$      & voltage threshold     & $0.8$      \\
						$x_{r}$, $y_{r}$     & voltage reset    & $-2.$   \\
						$\tau_m$ & membrane time constant      & $1.$        \\			
						\hline
						\multicolumn{2}{c}{\textbf{ Statistics of output spike trains 1 }} \\
						\hline
						$r_{1}$,$r_{2}$      & spikes per $\tau_m$   & $0.231$     \\
						$\CV^2_1$, $\CV^2_2$      &   squared coefficient of variation & $0.5$ \\			
						\hline
						\multicolumn{2}{c}{\textbf{ Model 2 (dark green) parameters}} \\
						\hline
						$x_{t}$, $y_{t}$      & voltage threshold     & $2.$  \\
						$x_{r}$, $y_{r}$     & voltage reset    & $-1.$   \\
						$\tau_m$ & membrane time constant      & $1.$  \\		
						\hline
						\multicolumn{2}{c}{\textbf{ Statistics of output spike trains 2 }} \\
						\hline
						$r_{1}$,$r_{2}$      & spikes per $\tau_m$   & $0.017$     \\
						$\CV^2_1$, $\CV^2_2$      &   squared coefficient of variation & $0.98$
					\end{tabular}}{}
				\end{table}
					
				\begin{table}[h!]
					\textbf{\refstepcounter{table}\label{sup_Tab:03} Table \arabic{table}.}{ Parameters for Fig.~~\ref{fig5}b, Fig.~\ref{fig5}c}				
					{\begin{tabular}{llr}			
							\hline
							\multicolumn{2}{c}{\textbf{ Neuron 1  parameters}} \\
							\hline
							Symbol    & Description & Value  \\
							\hline
							$x_{t}$    & voltage threshold  & in $[1.,0.5]$  \\
							$x_{r}$    & voltage reset    & in $[-2.5,-1.25]$ \\
							$\tau_m$ & membrane time constant      & $1.$ \\	
							\hline
							\multicolumn{2}{c}{\textbf{ Neuron 2 parameters}} \\
							\hline
							$y_{t}$     & voltage threshold     & in $[1.,0.5]$ \\
							$y_{r}$    & voltage reset    & in $[-2.5,-1.25]$  \\
							$\tau_m$ & membrane time constant      & $1.$ \\					
							\hline
							\multicolumn{2}{c}{\textbf{Reference parameters }} \\
							\hline
							$x_{t,\mathrm{ref}}$     & voltage threshold     &  $0.8$      \\
							$x_{r,\mathrm{ref}}$    & voltage reset    & $-2.$  \\
							$\tau_m$ & membrane time constant      & $1.$        \\	
							& for $x=\frac{V-\mu\tau}{\alpha\sigma\sqrt{\tau_m}}$   & $\alpha$ in $[0.8,1.5]$ , $\Delta\alpha=0.1$   \\ 
							\hline
							\multicolumn{2}{c}{\textbf{ $\sigma/\sigma_{ref}$ vs $C_{out}$ }} \\
							\hline
							$C_{in}$ & range of input correlations   & $[0,0.95]$\\
							$\Delta C_{in}$ & step of input correlations  & $0.05$
						\end{tabular}}{}
					\end{table}

					\begin{table}[h!]
						\textbf{\refstepcounter{table}\label{sup_Tab:04} Table \arabic{table}.}{ Parameters for Fig.~\ref{fig5}d }
						{\begin{tabular}{llr}
								\hline
								\multicolumn{2}{c}{\textbf{ Neuron 1  parameters}} \\
								\hline
								Symbol    & Description & Value  \\
								\hline
								$x_{t}$     & voltage threshold     &  $0$      \\
								$x_{r}$    & voltage reset    & $-2.5$   \\
								$\tau_m$ & membrane time constant      & $1.$        \\	
								\hline
								\multicolumn{2}{c}{\textbf{ Neuron 2 parameters}} \\
								\hline
								$y_{t}$     & voltage threshold     & $0.83$      \\
								$y_{r}$    & voltage reset    & $-1.66$  \\
								$\tau_m$ & membrane time constant      & $1.$        \\		
								\hline
								\multicolumn{2}{c}{\textbf{Input correlations }} \\
								\hline
								$C_\mathrm{in}$ & input correlation  & $0.9$
							\end{tabular}}{}
						\end{table}
						
						\begin{table}[h!]
							\textbf{\refstepcounter{table}\label{sup_Tab:05} Table \arabic{table}.}{ Parameters for Fig.~\ref{fig5}e }	
							
							{\begin{tabular}{llr}
									\hline
									\multicolumn{2}{c}{\textbf{ Neuron 1  parameters}} \\
									\hline
									Symbol    & Description & Value  \\
									\hline
									$x_{t}$     & voltage threshold     &  $1.$      \\
									$x_{r}$    & voltage reset    & $-2.5$   \\
									$\tau_m$ & membrane time constant      & $1.5$        \\	
									\hline
									\multicolumn{2}{c}{\textbf{ Neuron 2 parameters}} \\
									\hline
									$y_{t}$     & voltage threshold     & $0.5$      \\
									$y_{r}$    & voltage reset    & $-1.25$  \\
									$\tau_m$ & membrane time constant      & $1.$        \\		
									\hline
									\multicolumn{2}{c}{\textbf{Input correlations }} \\
									\hline
									$C_\mathrm{in}$ & input correlation  & $0.9$
								\end{tabular}}{}
							\end{table}
\clearpage						
\bibliographystyle{apsrev4-1}
\bibliography{biblio}

\end{document}